\documentclass[twocolumn,floatfix,superscriptaddress,showpacs,prb]{revtex4}
\usepackage{graphicx}
\usepackage{amsmath}
\usepackage{bm}
\begin{document}

\title{
Theory of the valley-valve effect in graphene nanoribbons}
\author{A. R. Akhmerov}
\affiliation{Instituut-Lorentz, Universiteit Leiden, P.O. Box 9506, 2300 RA Leiden, The Netherlands}
\author{J. H. Bardarson}
\affiliation{Instituut-Lorentz, Universiteit Leiden, P.O. Box 9506, 2300 RA Leiden, The Netherlands}
\author{A. Rycerz}
\affiliation{Institut f\"{u}r Theoretische Physik, Universit\"{a}t Regensburg, D-93040, Germany}
\affiliation{Marian Smoluchowski Institute of Physics, Jagiellonian University, Reymonta 4, 30--059 Krak\'{o}w, Poland}
\author{C. W. J. Beenakker}
\affiliation{Instituut-Lorentz, Universiteit Leiden, P.O. Box 9506, 2300 RA Leiden, The Netherlands}
\date{March 2008}
\begin{abstract}
A potential step in a graphene nanoribbon with zigzag edges is shown to be an intrinsic source of intervalley scattering --- no matter how smooth the step is on the scale of the lattice constant $a$. The valleys are coupled by a pair of localized states at the opposite edges, which act as an attractor/repellor for edge states propagating in valley $K/K'$. The relative displacement $\Delta$ along the ribbon of the localized states determines the conductance $G$. Our result $G=(e^{2}/h)[1-\cos(N\pi+2\pi\Delta/3a)]$ explains why the ``valley-valve'' effect (the blocking of the current by a \textit{p-n} junction) depends on the parity of the number $N$ of carbon atoms across the ribbon.
\end{abstract}
\maketitle

\section{Introduction}
\label{intro}

The massless conduction electrons in a two-dimensional carbon lattice respond differently to an electric field than ordinary massive electrons do. Because the magnitude $v$ of the velocity of a massless particle is independent of its energy, a massless electron moving along the field lines cannot be backscattered --- since that would require $v=0$ at the turning point. The absence of backscattering was discovered in carbon nanotubes,\cite{And98} where it is responsible for the high conductivity in the presence of disorder.

A graphene nanoribbon is essentially a carbon nanotube that is cut open along the axis and flattened. One distinguishes armchair and zigzag nanotubes, depending on whether the cut runs parallel or perpendicular to the carbon-carbon bonds. The edges of the nanoribbon fundamentally modify the ability of an electric field to backscatter electrons. As discovered in computer simulations by Wakabayashi and Aoki,\cite{Wak02} a potential step in a zigzag nanoribbon blocks the current when it crosses the Fermi level, forming a \textit{p-n} junction (= a junction of states in conduction and valence band). The current blocking was interpreted in Ref.\ \onlinecite{Ryc07} by analogy with the spin-valve effect in ferromagnetic junctions.\cite{Wol01} In this analogy the valley polarization in a zigzag nanoribbon plays the role of the spin polarization in a ferromagnet --- hence the name ``valley-valve'' effect.

\begin{figure}[tb]
\centerline{\includegraphics[width=0.8\linewidth]{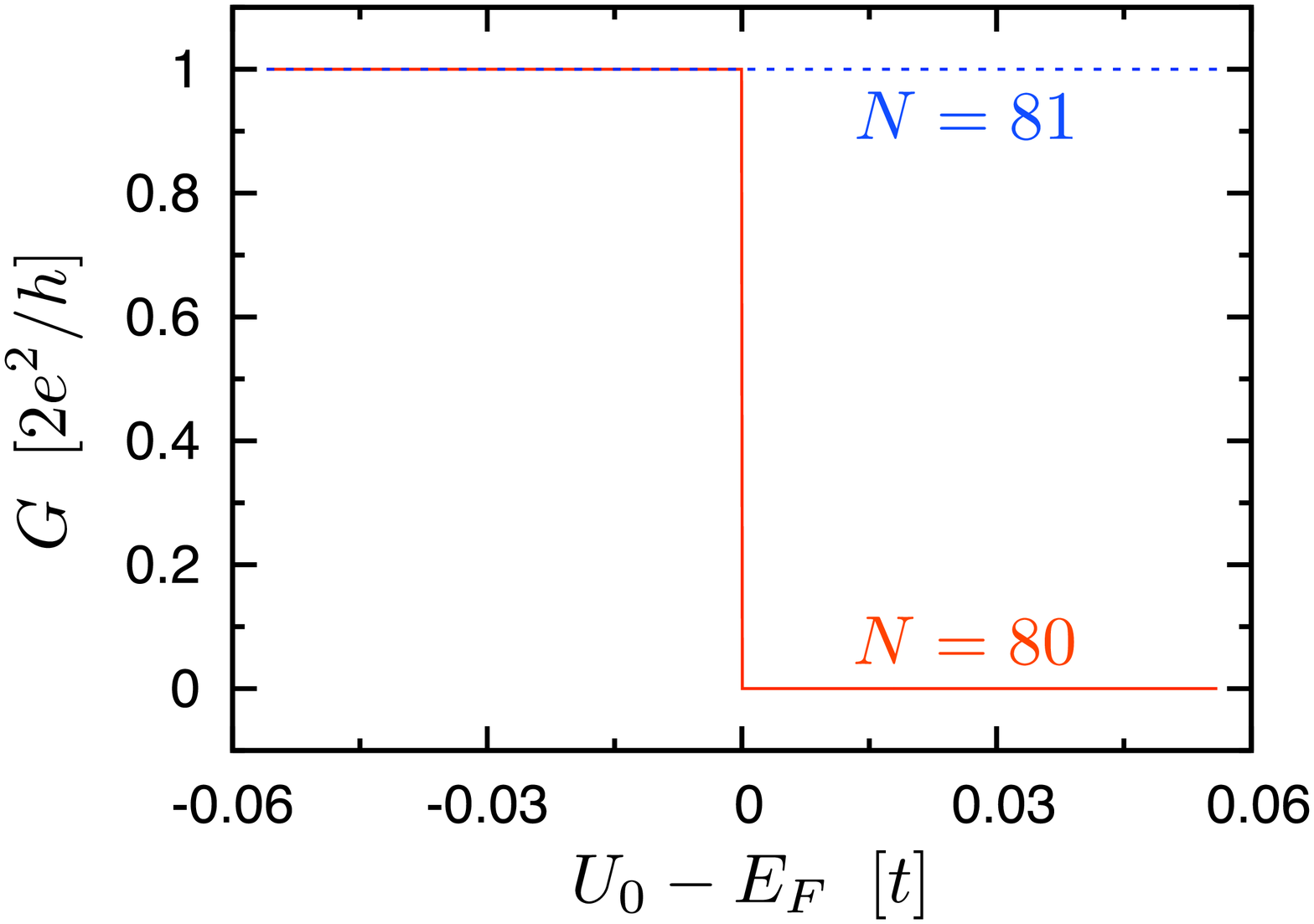}}\bigskip

\centerline{\includegraphics[width=0.8\linewidth]{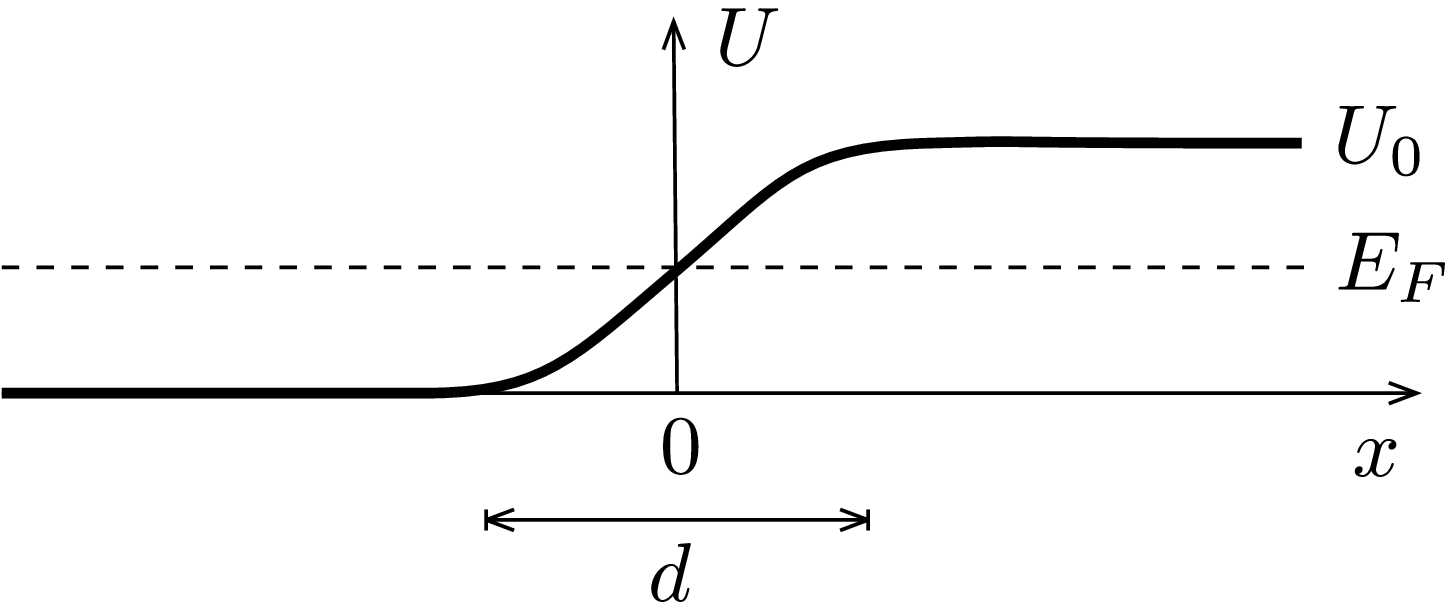}}
\caption{\label{fig_blocking}
Conductance $G$ of a zigzag nanoribbon containing a potential step $U=\tfrac{1}{2}U_{0}[1+\tanh(2x/d)]$. The red and blue curves are obtained by computer simulation of the tight-binding model of graphene, with parameters $d=10\,a$, $E_{F}=0.056\,t$, where $a$ is the lattice constant and $t$ is the nearest-neighbor hopping energy. Upon varying $U_{0}$ the conductance switches abruptly to zero when the Fermi level $E_{F}$ is crossed and a \textit{p-n} junction is formed (red solid curve; the deviation from an ideally quantized step function is $\lesssim 10^{-7}$). This ``valley-valve'' effect occurs only for an even number $N$ of carbon atom rows (zigzag configuration). When $N$ is odd (anti-zigzag configuration), the conductance remains fixed at $2e^{2}/h$ (blue dotted curve, again quantized within $10^{-7}$).
}
\end{figure}

\begin{figure}[tb]
\centerline{\includegraphics[width=0.9\linewidth]{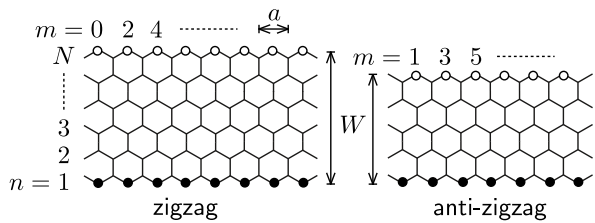}}
\caption{\label{fig_zigzag}
Nanoribbons in the zigzag configuration (left panel, $N$ even) and in the anti-zigzag configuration (right panel, $N$ odd). In both cases the atoms at opposite edges belong to different sublattices (indicated by black and white dots).
}
\end{figure}

It is the purpose of this paper to present a theory for this unusual phenomenon. A theory is urgently needed, because the analogy between spin valve and valley valve fails dramatically to explain the computer simulations of Fig.\ \ref{fig_blocking}: The current blocking by the \textit{p-n} junction turns out to depend on the parity of the number $N$ of atom rows in the ribbon. The current is blocked when $N$ is even (zigzag configuration), while it is not blocked when $N$ is odd (anti-zigzag configuration, see Fig.\ \ref{fig_zigzag}). This even-odd difference (first noticed in connection with the quantum Hall effect\cite{Two07}) is puzzling since zigzag and anti-zigzag nanoribbons are indistinguishable at the level of the Dirac equation,\cite{Bre06,note1} which is the wave equation that governs the low-energy dynamics in graphene.

\section{Breakdown of the Dirac equation at a potential step}
\label{breakdown}

\begin{figure}[tb]
\centerline{\includegraphics[width=0.8\linewidth]{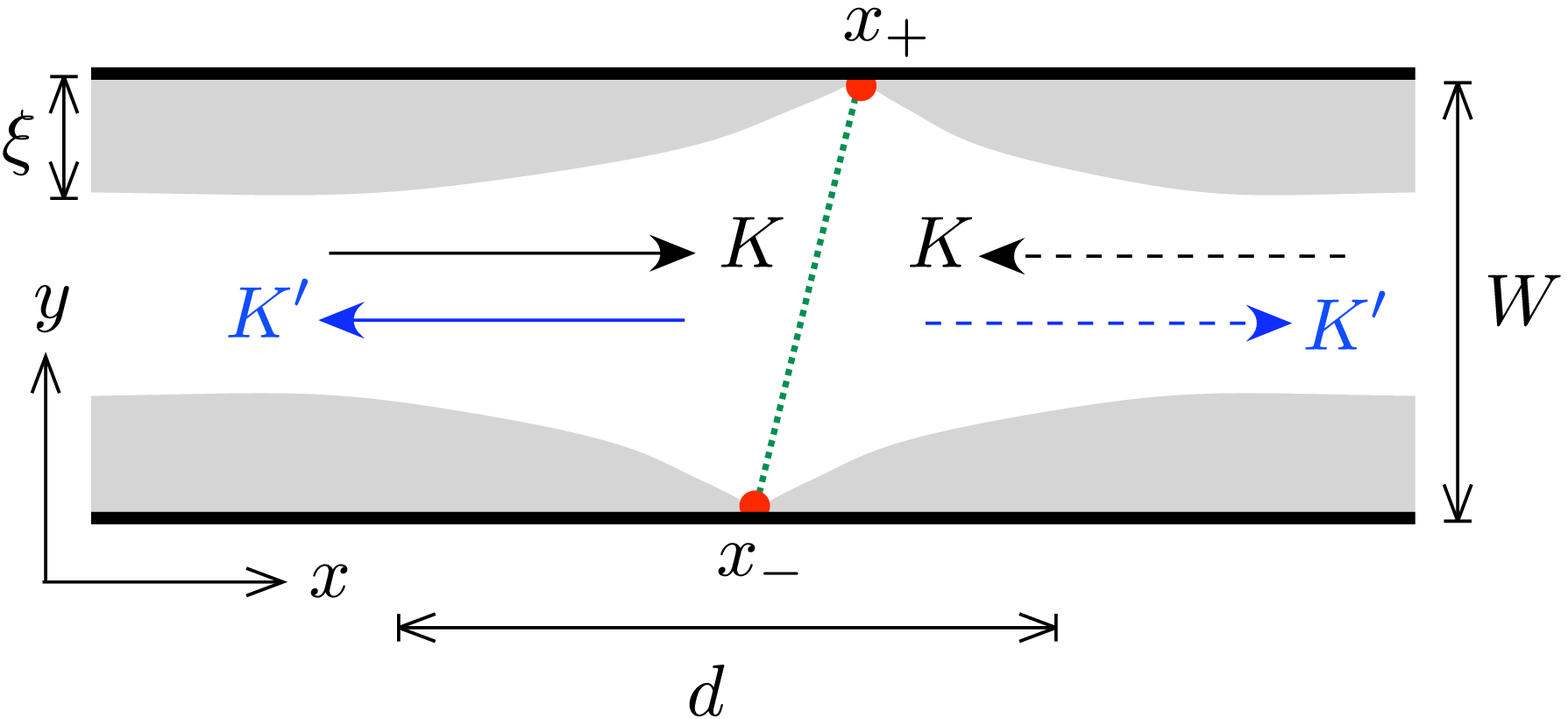}}

\centerline{\includegraphics[width=0.7\linewidth]{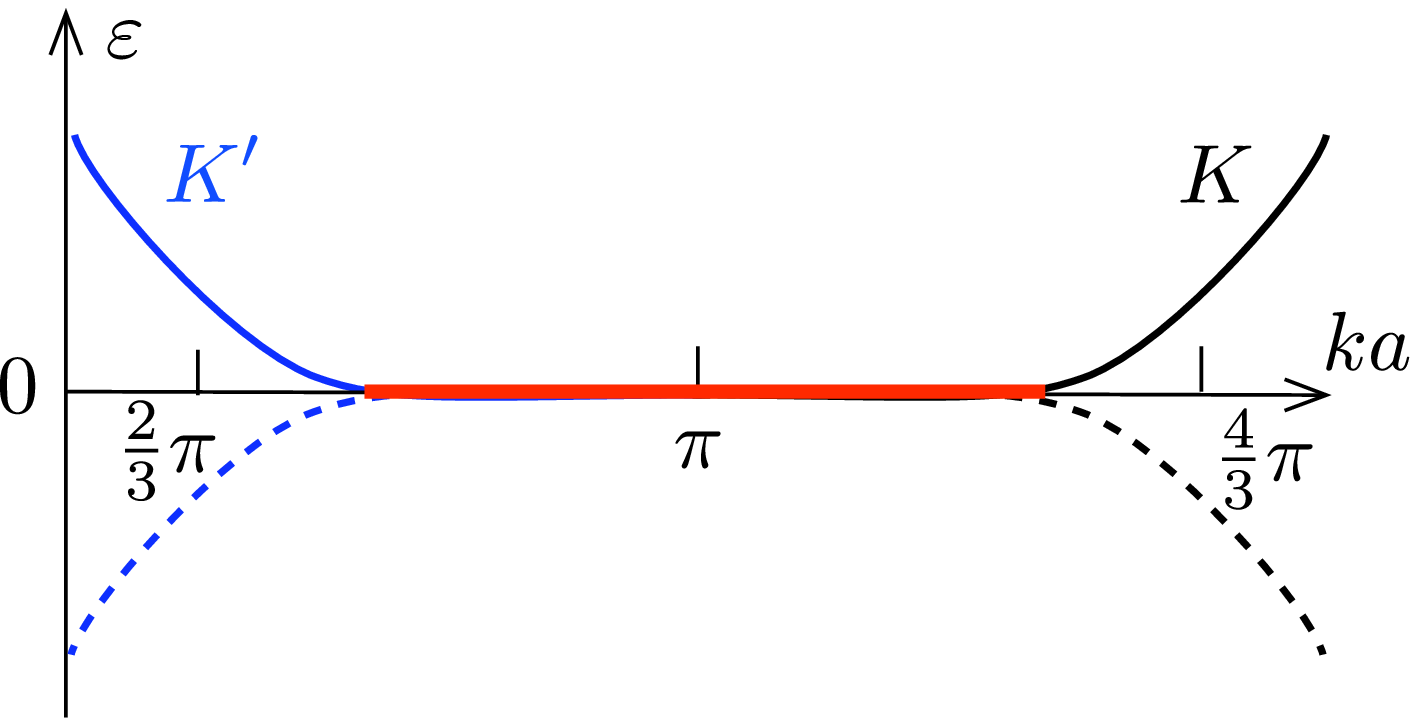}}
\caption{\label{fig_edge}
Top panel: Zigzag nanoribbon containing a \textit{p-n} interface from $x_{-}$ to $x_{+}$ (dotted line). The spatial extension of edge states in the lowest mode is indicated by the grey areas. Incoming edge states are in valley $K$, while outgoing edge states are in valley $K'$. The arrows indicate the direction of propagation, in the conduction band (solid) and valence band (dashed). The corresponding dispersion relations are plotted in the lower panel. The localized (dispersionless) edge state, responsible for the intervalley scattering, is indicated in red.
}
\end{figure}

The applicability of the Dirac equation rests on the assumption that a smooth potential step causes no intervalley scattering. As we now show, it is this assumption which fails in the \textit{p-n} junction, breaking the analogy between spin valve and valley valve. In the spin valve, a spin-up electron incident on a ferromagnetic junction which only transmits spin-down is simply reflected as spin-up. The current blocking can therefore be understood without invoking spin-flip scattering. In the valley valve, however, an electron in valley $K$ incident on a \textit{p-n} junction which only transmits valley $K'$ cannot be reflected in valley $K$. Both transmission and reflection require a switch of the valley from $K$ to $K'$ (see Fig.\ \ref{fig_edge}). We conclude that {\em a \textit{p-n} junction in a zigzag nanoribbon is an intrinsic source of intervalley scattering}. It does not matter how smooth the potential step might be, since the incoming and outgoing states are from different valleys, the scattering must switch valleys to preserve the current.

As we have illustrated in Fig.\ \ref{fig_edge}, the source of intervalley scattering is a pair of localized edge states at the \textit{p-n} interface. It is well-known that the lowest mode in a zigzag nanoribbon is confined near the edges.\cite{Nak96} The transverse extension $\xi(\varepsilon)\sim W/\ln|\varepsilon W/\hbar v|$ of an edge state depends on the kinetic energy $\varepsilon=E_{F}-U$. We define the \textit{p-n} interface as the line where $E_{F}-U(x,y)=0$. This line intersects the two edges at the points $\bm{r}_{\pm}=(x_{\pm},y_{\pm})$, with $y_{+}=(\tfrac{3}{2}N-1)a/\sqrt{3}\equiv W$ and $y_{-}=0$ the $y$-coordinates of the row of atoms at the upper and lower edge, respectively. (Note that $\bm{r}_{\pm}$ is well-defined also for a smooth interface.) Upon approaching the \textit{p-n} interface, $\varepsilon$ decreases from $E_{F}$ to $0$ hence $\xi$ decreases from $\xi(E_{F})\equiv\xi_{0}$ to a minimal value of order of the lattice constant $a$. An electron incident on the \textit{p-n} junction in valley $K$ is therefore attracted to a pair of localized edge states centered at $\bm{r}_{\pm}$. Their wave vector $k$ spans the interval of order $1/a$ between the valleys $K$ and $K'$ --- thereby allowing for the intervalley scattering needed to repel the electron into valley $K'$.

\section{Scattering theory beyond the Dirac equation}

Now that we have identified the mechanism for intervalley scattering, we need to calculate the coupling of the propagating edge states to the pair of localized edge states, in order to determine whether an incident electron is transmitted or reflected at the \textit{p-n} interface. For this purpose we have developed a scattering theory based directly on the tight-binding Hamiltonian, 
\begin{equation}
H_{0}=t\sum_{\rm neighbors}|n,m\rangle\langle n',m'|,\label{H0def}
\end{equation}
thereby going beyond the Dirac equation. The calculation is outlined below, but we first present the result --- which is remarkably simple: The transmission probability $T$ [and hence the conductance $G=(2e^{2}/h)T$] is determined by the lateral displacement $\Delta=x_{+}-x_{-}$ of the localized states, according to
\begin{equation}
T=\tfrac{1}{2}-\tfrac{1}{2}\cos(N\pi+2\pi\Delta/3a),\label{Tresult}
\end{equation}
for $W\gg\Delta$. This is the central result of our paper.

We have derived Eq.\ \eqref{Tresult} by projecting the tight-binding Hamiltonian onto the pair of (nearly degenerate) lowest modes, and then solving a scattering problem in $k$-space. As illustrated in Fig.\ \ref{fig_edge} (lower panel), incoming and outgoing states have wave vectors near $k_{\rm in}\approx 4\pi/3a$ and $k_{\rm out}\approx 2\pi/3a$, respectively. The unitary transformation of an incoming state into an outgoing state is governed by the $2\times 2$ transfer matrix $M$ in the linear relation
\begin{equation}
\Psi(k)=M(k,k')\Psi(k').\label{Mdef}
\end{equation}
Here we have introduced the two-component wave function $\Psi(k)=(\psi_{k}^{+},\psi_{k}^{-})$ in $k$-space. (For later use we also introduce the Pauli matrices $\sigma_{i}$ acting on the $\pm$ degree of freedom of the nearly degenerate lowest modes, with $\sigma_{0}$ the $2\times 2$ unit matrix.) Once we know $M$, the scattering matrix $S=\Omega_{\rm out} M(k_{\rm out},k_{\rm in})\Omega^{\dagger}_{\rm in}$ follows by a change of basis such that $\Omega_{X}\Psi(k_{X})$ (with $X$ labeling ``in'' or ``out'') has the first component in the conduction band (left end of the nanoribbon) and the second component in the valence band (right end of the nanoribbon).

An analytical calculation is possible for $W\gg\xi_{0}$, when we can approximate the lowest modes $\psi_{k}^{\pm}=2^{-1/2}(\psi_{k}^{A}\pm\psi_{k}^{B})$ by \cite{Nak96}
\begin{align}
\psi_{k}^{A}(m,n)={}&C(k)e^{imka/2}[-2\cos(ka/2)]^{n-1}\pi_{n+m+1},\label{PsikAdef}\\
\psi_{k}^{B}(m,n)={}&C(k)e^{imka/2}[-2\cos(ka/2)]^{N-n}\pi_{n+m},\label{PsikBdef}
\end{align}
with $C(k)=\sqrt{-1-2\cos ka}$ a normalization factor. We have defined $\pi_{p}=1$ if $p$ even and $\pi_{p}=0$ if $p$ odd. The integer $n$ labels the row of atoms in the $y$-direction and $m$ labels the column of atoms in the $x$-direction (see Fig.\ \ref{fig_zigzag}). This approximation is accurate in the whole range $(2\pi/3a,4\pi/3a)$ of $k$, except within an interval of order $1/W$ from the end points.
 The wave functions $\psi_{k}^{A}$, $\psi_{k}^{B}$ are edge states, extended either along the lower edge (on the $A$ sublattice, indicated by black dots in Fig.\ \ref{fig_zigzag}) or along the upper edge (on the $B$ sublattice, white dots).
 
The nearest-neigbor tight-binding Hamiltonian \eqref{H0def} is diagonal in the basis of the modes $\psi_{k}^{\pm}$, with matrix elements
\begin{align}
&\langle k,\pm|H_{0}|k',\pm\rangle=\pm\varepsilon(k)a^{-1}\delta(k-k'),\label{H0matrix}\\
&\varepsilon(k)=2tC(k)^{2}[-2\cos(ka/2)]^{N}. \label{epsilondef}
\end{align}
Since $\varepsilon(\pi/a-\delta k)=(-1)^{N}\varepsilon(\pi/a+\delta k)$, the parity of $N$ determines whether or not $\psi_{k}^{\pm}$ switches between conduction and valence band as $k$ crosses the point $\pi/a$. This band switch is at the origin of the parity dependence of the valley-valve effect, since it introduces a parity dependence of the matrices $\Omega_{\rm in}=\sigma_{0}$, $\Omega_{\rm out}=\sigma_{1}^{N}$ that transform the transfer matrix into the scattering matrix.

We model the \textit{p-n} interface by a linear potential profile,
\begin{equation}
U_{nm}=U_{x}m/2+U_{y}(n-N/2),\label{Udef}
\end{equation}
tilted by an angle $\theta=\arctan(\frac{2}{3}\sqrt{3}\,U_{y}/U_{x})$. Upon projection onto the two-component space spanned by $\Psi(k)$, the Hamiltonian $H=H_{0}+U$ becomes an integral kernel $H(k,k')$ with a $2\times 2$ matrix structure:
\begin{align}
H(k,k')={}&\varepsilon(k)a^{-1}\delta(k-k')\sigma_{3}+iU_{x}
a^{-2}\frac{d}{dk}\delta(k-k')\sigma_{0}\nonumber\\
&+\tfrac{1}{2}Na^{-1}U_{y}\delta(k-k')\sigma_{1}.\label{Ulowest}
\end{align}
The integral equation 
\begin{equation}
a\int^{4\pi/3a}_{2\pi/3a} dk'\,H(k,k')\Psi(k')=E\Psi(k)
\end{equation}
 amounts to a system of two first order differential equations:
\begin{equation}
\left(\varepsilon(k)\sigma_3 + i U_x a^{-1} \sigma_0\frac{d}{d k}+\frac{1}{2}N U_y\sigma_1\right)\Psi(k)=E \Psi(k).
\end{equation}
This system gives directly an expression for the transfer matrix,
\begin{gather}
M(k,k')=\exp\left[i(k'-k)a E/U_x\right]{\cal T}\exp\left[i\int_{k'}^{k}dq\,\Omega(q)\right],\label{Psik1k2}\\
\Omega(q)=\frac{\varepsilon(q)a}{U_{x}}\sigma_{3}+\frac{\Delta}{2}\sigma_{1}.
\end{gather}
The scalar phase factor $\exp[i(k'-k)a E/U_{x}]$ has no effect on the transmission probability, so we will omit it in what follows.
The symbol ${\cal T}$ orders the operators in the exponent with respect to the variable $q$ (from $q=k$ at the left to $q=k'$ at the right). The scattering matrix follows from
\begin{equation}
S=\sigma_{1}^{N}M(k_{\rm out},k_{\rm in}).\label{SMrelation}
\end{equation}

We may evaluate Eq.\ \eqref{Psik1k2} analytically if $W\gg\Delta$, because then the integration interval can be separated into subintervals in
which the contribution of one of the terms can be neglected. The calculation is described in App.\ \ref{appA}. The result is
\begin{equation}
M(k_{\rm out},k_{\rm in})= e^{i\alpha\sigma_{3}}\exp[-i(\pi\Delta/3a)\sigma_{1}]e^{i\alpha'\sigma_{3}},\label{Mresult}
\end{equation}
with a phase shift $\alpha=(-1)^{N}\alpha'$ that need not be determined. Substitution into Eq.\ \eqref{SMrelation} yields the result \eqref{Tresult} for the transmission probability $T=|S_{12}|^{2}$.

The regime $\Delta\gtrsim W$ can be analyzed by a numerical evaluation of the integral \eqref{Psik1k2}. The result, shown in Fig.\ \ref{fig_decay} (solid curve), is that the conductance oscillations are damped for $\Delta\gtrsim W$.

\section{Comparison with computer simulations}
\label{comparison}

\begin{figure}[tb]
\centerline{\includegraphics[width=0.8\linewidth]{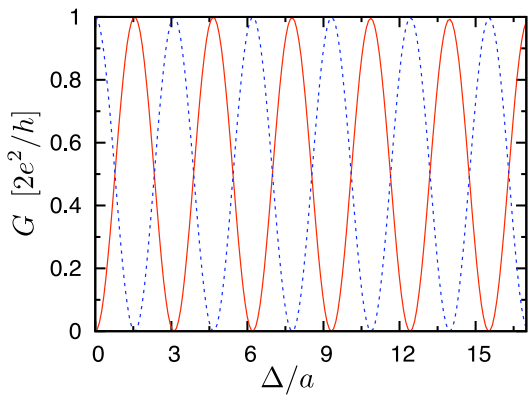}}
\caption{\label{fig_tilted}
Conductance for a tilted \textit{p-n} interface, with potential $U=\tfrac{1}{2}U_{0}\{1+\tanh[2(x-y\tan\theta)/d]\}$, at fixed $U_{0}=2E_{F}$ as a function of the relative displacement $\Delta\equiv x_{+}-x_{-}=W\tan\theta$ of the intersection of the interface with the edges of the nanoribbon. The parameters are the same as in Fig.\ \ref{fig_blocking}, which corresponds to $\Delta=0$. The data from this computer simulation is described by the analytical result \eqref{Tresult}.
}
\end{figure}

The current blocking ($T=0$) obtained in the computer simulations of Refs.\ \onlinecite{Wak02,Ryc07} is the special case $N$ even, $\Delta=0$, corresponding to a zigzag configuration with potential $U$ independent of $y$. In the anti-zigzag configuration ($N$ odd) we have instead $T=1$, in accord with the simulations of Fig.\ \ref{fig_blocking}. More generally, we can tilt the interface so that $\Delta\neq 0$. The simulations for a tilted \textit{p-n} interface shown in Fig.\ \ref{fig_tilted} are well described by the analytical result \eqref{Tresult}, for $\Delta\ll W\simeq 70\,a$. Note in particular the sum rule $G(N)+G(N+1)\approx e^{2}/h$, first observed in the computer simulations of Ref.\ \onlinecite{Two07}.

\begin{figure}[tb]
\centerline{\includegraphics[width=0.8\linewidth]{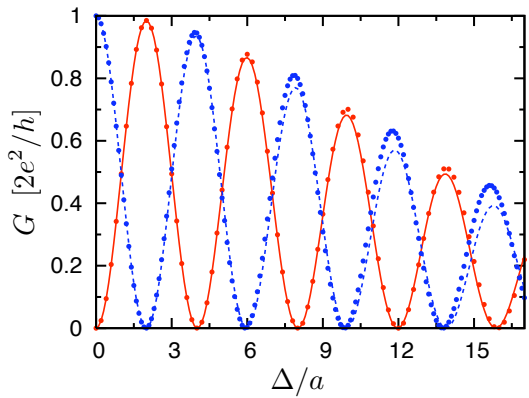}}
\caption{\label{fig_decay} Comparison between results of computer simulations (dots) and numerical evaluation of Eq.\ \eqref{Psik1k2}. The parameters are $N=20$ (solid line) and $N=21$ (dashed line), $U_0=2 E_{F}=0.0058\,t$ and $d\equiv E_{F}/U_{x}=100\,a$.
}
\end{figure}

For larger $\Delta/W$ a phase shift appears and a reduction of the amplitude of the oscillations, with $G\simeq 0$ for $\Delta\gtrsim W$. We compare the conductance calculated by numerical evaluation of the integral \eqref{Psik1k2} with the data from the computer simulations and find good agreement, see Fig.\ \ref{fig_decay}.

\section{Extensions of the theory}
\label{extensions}

The theory presented so far can be extended in several ways.

We have assumed that the width $W$ of the nanoribbon is sufficiently narrow that there is only a single propagating mode at the Fermi level, which requires\cite{Ryc07} $W<4a\tau/E_{F}$. The assumption can be relaxed in the case of a smooth \textit{p-n} interface, because higher modes have an exponentially small transmission probability if the Fermi wavelength\cite{Che06} $\lambda_F\simeq W\ll d$.

Next-nearest-neighbor hopping was not included in the theory, and one might be concerned that it could modify our result substantially because the edge states are then no longer dispersionless.\cite{Per06} We have found that this is actually not a relevant perturbation: Next-nearest-neigbor hopping (with hopping energy $t'$) adds a term $2t'(2+\cos ka)a^{-1}\delta(k-k')\sigma_{0}$ to the projected Hamiltonian \eqref{Ulowest}. This is an irrelevant perturbation because its only effect is to multiply the transfer matrix \eqref{Psik1k2} by a scalar phase factor.

\begin{figure}[tb]
\centerline{\includegraphics[width=0.8\linewidth]{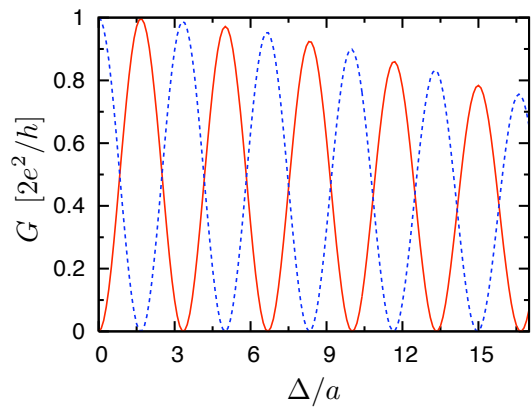}}
\caption{\label{fig_nnh} The conductance of a tilted \textit{p-n} interface with next-nearest neighbor hopping included ($t^\prime/t = 0.1$). The parameters of the ribbon are $E_F = 0.19\,t$, $U_0=0.16\,t$, and $d = 100\,a$. The number of atoms across the ribbon is $N=40$ (solid line) and $N=41$ (dashed line).
}
\end{figure}

As a check, we have repeated the computer simulations with the inclusion of next-nearest-neighbor hopping \cite{note4} in the tight-binding model (for the realistic ratio $t'/t=0.1$). As shown in Fig.\ \ref{fig_nnh}, the result \eqref{Tresult} still applies for $\Delta\ll W$.

Eq.\ \eqref{Tresult} was derived for a linear potential profile $U$, but the derivation can be extended to include a smoothly varying potential landscape $\delta U$ (smooth on the scale of the lattice constant). Electrostatic disorder therefore affects the conductance only through the lateral displacement $\Delta$ of the points on the boundary at which $U+\delta U-E_{F}=0$. 

Edge disorder cannot be accounted for in this simple way, but in view of the small lateral extension of the localized edge state we might not need a well-defined zigzag edge over long distances in order for Eq.\ \eqref{Tresult} to apply.

\section{Conclusion}
\label{conclude}
 
In conclusion, we have presented a theory for the current blocking by a \textit{p-n} junction in a zigzag nanoribbon. The dependence on the parity of the number $N$ of atoms across the ribbon, not noticed in earlier computer simulations,\cite{Wak02,Ryc07} is explained in terms of the parity of the lowest mode under a switch of sublattice: Incident and transmitted modes have opposite parity for $N$ even, leading to complete reflection ($G=0$), while they have the same parity for $N$ odd, leading to complete transmission ($G=2e^{2}/h$). A variation of the electrostatic potential in the direction transverse to the ribbon can invert the parity dependence of the conductance, while preserving the sum rule $G(N)+G(N+1)\approx 2e^{2}/h$. 

This switching behavior may have device applications, if the structure of the edges can be controlled (which is not the case in presently available samples). Even if such control is not forthcoming, the mechanism for current blocking proposed here can be operative in an uncontrolled way in disordered nanoribbons, producing highly resistive \textit{p-n} interfaces at random positions along the ribbon. Conduction through the resulting series of weakly coupled regions would show an activated temperature dependence as a result of the Coulomb blockade,\cite{Sol07} as observed experimentally.\cite{Han07,Che07}

\acknowledgments
This research was supported by the Dutch Science Foundation NWO/FOM. AR acknowledges support from the Alexander von Humboldt-Stiftung,
the Polish Ministry of Science (Grant No.\ 1--P03B--001--29), and the Polish Science Foundation (FNP). We acknowledge the help of J. Tworzyd{\l}o with the computer simulations.

\appendix
\section{Evaluation of the transfer matrix}
\label{appA}

To evaluate the transfer matrix $M(k_{\rm{out}},k_{\rm{in}})$ in the regime $W\gg\Delta$ we use the fact that the energy $\varepsilon(k)$ of the lowest modes decays exponentially $\sim\exp(-Na\delta k)$ away from the integration limits $k_{\rm in},k_{\rm out}$ [see Eq.\ \eqref{epsilondef}]. We separate the integration in the momentum-ordered exponent \eqref{Psik1k2} into three intervals:
\begin{multline}
M(k_{\rm{out}},k_{\rm{in}})={\cal T} \exp\left[i\int_{k_2}^{k_{\rm{out}}}\Omega(q)dq\right] \\
{\cal T} \exp\left[i\int_{k_1}^{k_2}\Omega(q)dq\right]{\cal T} \exp\left[i\int_{k_{\rm{in}}}^{k_1}\Omega(q)dq\right].
\end{multline}
We choose $k_{\rm{in}}-k_1= k_2-k_{\rm{out}}\gtrsim 1/W$, such that $|\varepsilon(k_1)|=|\varepsilon(k_2)|\lesssim U_{x}$. Then the contribution of the term $\sigma_1\Delta/2$ in $\Omega(q)$ to the integrals over the first and the third intervals is of order $\Delta/W \ll 1$, so that this term may be neglected. The contribution of the term $\sigma_3\varepsilon(q)a/U_x$ to the integral over the second interval is of order $a/W\ll 1$ so it can also be neglected. The three integrals can now be evaluated analytically, with the result:
\begin{equation}
M(k_{\rm{out}},k_{\rm{in}})= e^{i \alpha \sigma_3}
\exp\left[i (k_2-k_1) \frac{\Delta}{2}\sigma_1\right]
e^{i \alpha' \sigma_3}.
\end{equation}
This is equivalent to Eq.\ \eqref{Mresult} since $k_2-k_1=-2\pi/3 a+{\cal O}(1/W)$.


\begin{thebibliography}{99}
\bibitem{note1} The dependence of boundary conditions on the number $N$ of atoms across the ribbon is a key distinction between zigzag and armchair edges. The boundary condition of the Dirac equation for an armchair nanoribbon depends on $N$ (modulo 3), but there is no $N$-dependence for a zigzag nanoribbon.
\end{thebibliography}
\end{document}